\documentclass[conference]{IEEEtran}
\IEEEoverridecommandlockouts
\usepackage{amsmath}
\usepackage{algorithmic} 
\usepackage{array} 
\usepackage{stfloats} 
\usepackage{url} 
\usepackage{graphicx} 
\usepackage[nocompress]{cite} 
\usepackage[nolist,nohyperlinks]{acronym} 
\usepackage{threeparttable} 
\usepackage{balance} 
\usepackage{multirow} 
\usepackage{xcolor}
\usepackage{flushend}
\usepackage{mathtools} 
\usepackage{wrapfig}
\ifCLASSOPTIONcompsoc
\usepackage[caption=false,font=footnotesize,labelfont=sf,textfont=sf]{subfig} 
\else
 \usepackage[caption=false,font=footnotesize]{subfig}
\fi

\usepackage[T1]{fontenc}
\usepackage{newtxtext}  

\usepackage{cite}
\usepackage{amsmath,amssymb,amsfonts}
\usepackage{algorithmic}
\usepackage{graphicx}
\usepackage{textcomp}
\usepackage{xcolor}
\usepackage{color, soul}
\def\BibTeX{{\rm B\kern-.05em{\sc i\kern-.025em b}\kern-.08em
    T\kern-.1667em\lower.7ex\hbox{E}\kern-.125emX}}

\begin{acronym}
\acro{SLAM}{Simultaneous Localization And Mapping}
\acro{BLE}{Bluetooth Low Energy}
\acro{UWB}{Ultra-Wideband}
\acro{NFC}{Near-field communication}
\acro{NLOS}{non-line-of-sight}
\acro{LOS}{line-of-sight}
\acro{EKF}{Extended Kalman Filter}
\acro{PDR}{Pedestrian Dead Reckoning}
\acro{RSS}{Received Signal Strength}
\acro{RTT}{Round Trip Time}
\acro{CNN}{convolutional neural network}
\acro{IoT}{Internet of Things}
\acro{IMU}{Inertial Measurement Unit}
\end{acronym}

\def \sys {\emph{UbiLoc}}
\usepackage{wrapfig}

\begin{document}

\title{AirTags for Human Localization, Not Just Objects}
\author{\IEEEauthorblockN{Mohamed I. Hany}
\IEEEauthorblockA{\textit{Dept of Computer Science \& Engineering} \\
\textit{The American University in Cairo}\\
Cairo, Egypt \\
mohamed.ihany@aucegypt.edu}
\and
\IEEEauthorblockN{Hamada Rizk}
\IEEEauthorblockA{\textit{Dept of Information Networking} \\
\textit{Osaka University, Japan}\\
\textit{Tanta University, Egypt}\\    
hamada\_rizk@ist.osaka-u.ac.jp}
\and
\IEEEauthorblockN{Moustafa Youssef}
\IEEEauthorblockA{\textit{Dept of Computer Science \& Engineering} \\
\textit{The American University in Cairo}\\
Cairo, Egypt \\
moustafa-youssef@aucegypt.edu}
}

\maketitle

\begin{abstract}
Indoor localization has become increasingly important due to its wide-ranging applications in indoor navigation, emergency services, the \ac{IoT}, and accessibility for individuals with special needs. Traditional localization systems often require extensive calibration to achieve high accuracy. We introduce \sys{}, an innovative, calibration-free indoor localization system that leverages Apple AirTags in a novel way to localize users instead of tracking objects. By utilizing the ubiquitous presence of AirTags and their \ac{UWB} technology, \sys{} achieves centimeter-level accuracy, surpassing traditional WiFi and \ac{BLE} systems. \sys{} addresses key challenges, including ranging errors caused by multipath and noise, through a novel AirTag selection technique. The system operates without the need for manual calibration, ensuring robustness and self-maintenance. Deployed on various Apple devices and tested in real-world environments, \sys{} achieved median localization errors as low as $26~cm$ in a campus building and $31.5~cm$ in an apartment setting. These results demonstrate that \sys{} is the first system to offer reliable, cm-level accuracy using widely available technology without requiring calibration, making it a promising solution for next-generation indoor localization systems.

\end{abstract}


\section{Introduction}
Indoor localization has become a critical research area due to its vast applications in indoor navigation, emergency services, and smart environments, as well as its importance for accessibility solutions for individuals with special needs \cite{want1992active, lu1997globally}. Achieving centimeter-level accuracy is crucial for these applications, as it ensures precise and reliable performance in complex indoor settings. However, current localization technologies such as WiFi and \ac{BLE} struggle to deliver the required accuracy without significant calibration efforts \cite{jiang2012ariel, park2010growing, shokry2017tale, wang2012NoNeedWardrive} \cite{youssef2005horus, rizk2022robust, rizk2021device}. WiFi-based systems, for instance, often depend on fingerprinting techniques, which require extensive data collection and suffer from high temporal and environmental variability \cite{bahl2000radar,youssef2005horus, rizk2019device}. This process not only introduces high labour costs but also degrades accuracy over time as environmental conditions change \cite{vo2015survey, ibrahim2018zero}. Similarly, BLE offers only coarse proximity-based tracking, which is insufficient for fine-grained applications. As indoor environments become more dynamic and complex, the need for centimeter-level accuracy, particularly for applications like assistive technologies and precise asset tracking, is growing rapidly. Therefore, there is a pressing demand for a ubiquitous, calibration-free system that can provide consistent cm-level indoor positioning.

\begin{figure}[!t]
\centerline{\includegraphics[width=0.8\linewidth]{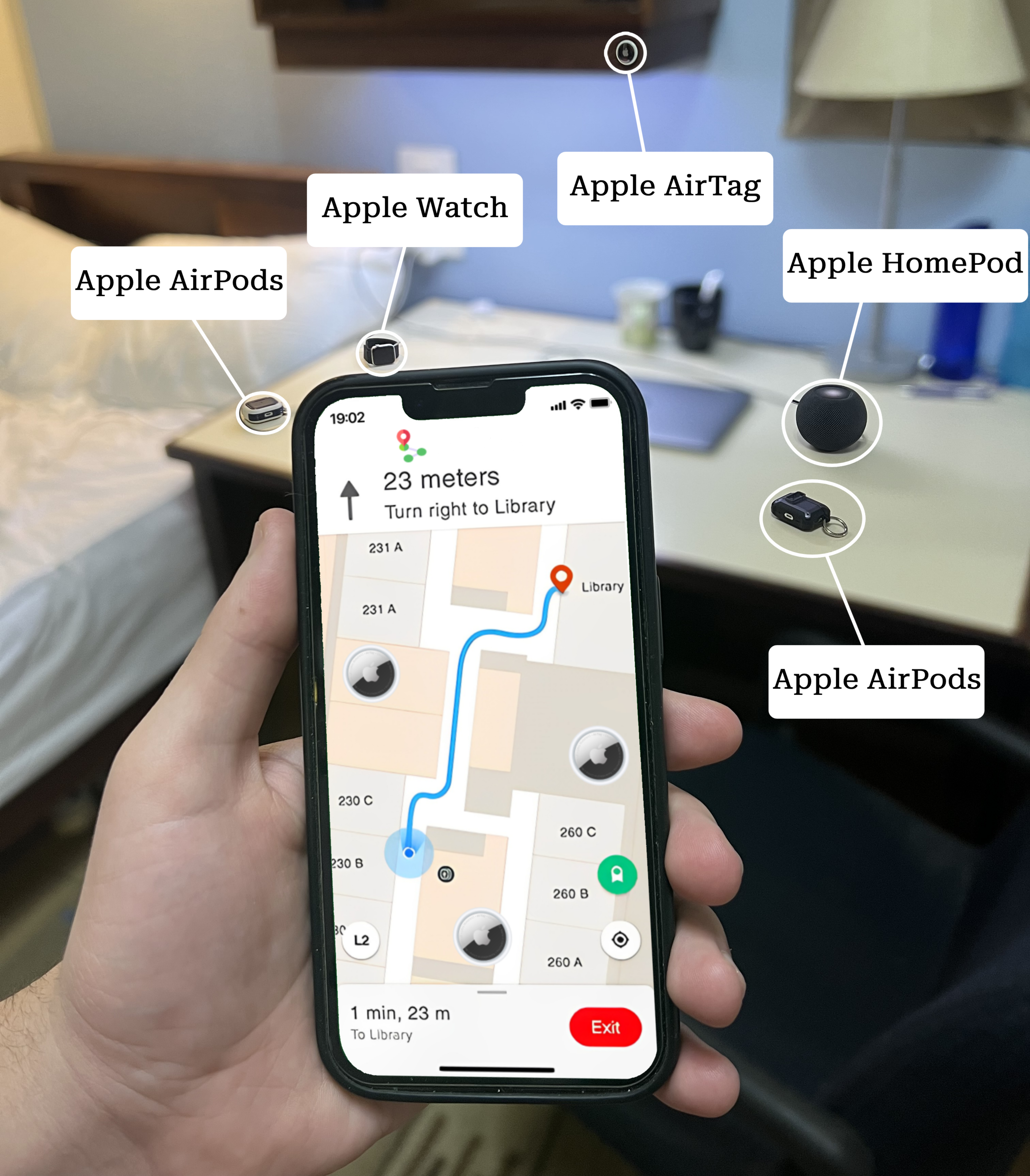}}
	\caption{An example of our futuristic vision for an indoor positioning system involves using Apple U1-chip products like AirPods, iWatch, HomePod, and AirTags. In this vision, \sys{} leverages detected AirTags (anchor points) to localize the user within the testbed, providing a seamless and precise indoor positioning solution.}
	\label{fig:future-vision}
  \vspace{-0.7cm}
\end{figure}

Apple AirTags are becoming increasingly ubiquitous, with the global market expected to grow from USD 3.88 billion in 2022 to USD 29.58 billion by 2031 \cite{Airtag_Website}. Equipped with \ac{UWB} technology, AirTags can provide highly accurate ranging measurements, enabling centimeter-level precision in distance estimation. \ac{UWB} offers substantial advantages over other wireless technologies such as WiFi and \ac{BLE}, particularly in terms of spatial resolution and resistance to interference. This makes AirTags a promising candidate for achieving cm-level indoor localization, opening new possibilities for precision tracking in environments like smart homes, offices, and healthcare facilities. However, despite this potential, the current studied use case for AirTags focuses exclusively on asset tracking, primarily for locating personal gadgets, rather than for localizing users themselves. This limitation highlights the untapped potential of AirTags in user-centric indoor localization applications.

In this paper, we present our proposed \sys{}, the first system to invert the typical usage of AirTags by leveraging them to localize the user instead of tracking personal gadgets. We achieve this in a completely calibration-free manner. The core idea is to utilize the ranging and angular measurements from AirTags placed in the environment to determine the user’s position through a multilateration-based approach. However, this approach presents several technical challenges. First, ranging errors are introduced due to multipath effects and noisy signal propagation. We address these errors by optimizing the selection of AirTags, refining the localization process and improving accuracy significantly. In Figure~\ref{fig:future-vision}, we present a futuristic vision for an indoor positioning system based on commercial-off-the-shelf. A network of AirTags (commercial-off-the-shelf devices) will be used to cover the whole layout plan. The AirTags network density will affect positioning accuracy.

To address these challenges, we employ a multilateration approach that uses the available AirTag locations, making the system both calibration-free and self-maintaining. This method eliminates the need for manual calibration by automatically adjusting to the environment. Additionally, we conducted a study to analyze the correlation between the user’s relative position to the AirTag and the resulting ranging error. The insights from this study allowed us to design a novel technique for selecting the most reliable AirTag measurements. By prioritizing these signals, we significantly improve localization accuracy and reduce the impact of noise and multipath effects.

We developed and deployed the proposed \sys{} on multiple Apple phones across two real-world test environments. Our experiments showed remarkable improvements in localization accuracy, achieving median errors as low as $26~cm$ in a campus building and $31.5~cm$ in an apartment. This establishes our system as the first to achieve cm-level accuracy using widely available technology without requiring any calibration. In summary, our main contributions are:
\begin{itemize}
    \item We present the first system to repurpose AirTags for providing cm-level accuracy in user localization.
    \item We identify and overcome several challenges related to the performance of \sys{}, including managing ranging errors due to multipath and optimizing the selection of AirTags.
    \item We implement, refine, and evaluate \sys{} in two distinct testbeds, demonstrating its potential and practical feasibility as a reliable indoor localization solution.
\end{itemize}

The rest of the paper is structured as follows: Section~\ref{sec:background-feasibility} presents a thorough feasibility analysis, emphasizing the key advantages of using AirTags for localization. Next, Section~\ref{sec:System-Overview} outlines the system design and addresses various practical challenges. In Section~\ref{sec:system-evaluation}, we assess the system’s performance and benchmark it against current state-of-the-art solutions. Finally, Sections~\ref{sec:Related-Work} and \ref{sec:conclusion} provide an overview of related work and offer concluding remarks, respectively.

\section{Background and Feasibility Study}\label{sec:background-feasibility}
This section begins with an in-depth introduction to AirTags, covering key background information. Following that, we perform a feasibility analysis highlighting the advantages of using AirTags for indoor localization.

\subsection{AirTag Tearing Down}\label{sec:airtag-teardown}
First, we want to tear down the Apple AirTag to understand its internal components and how they contribute to indoor localization. AirTags are small, coin-sized devices equipped with several key technologies, including a Bluetooth Low Energy \ac{BLE} chip and an \ac{UWB} chip, both crucial for communication and precise positioning \cite{heinrich2023smartphones, throughfloor_vsvecova2019through, song2019uwb, zhao2021uloc}. \ac{UWB} chip enables accurate time-of-flight measurements, which are essential for providing both ranging and angular readings, allowing for more precise localization. In addition, the AirTag contains an accelerometer, which helps detect movement. By understanding the roles of these components, we can better optimize AirTags for our proposed system, shifting their use from simple asset tracking to accurate user localization. This teardown helps us maximize the potential of AirTag technology for our application in indoor positioning.

\begin{figure}[!t]
    \begin{minipage}[b]{1\linewidth}
    \centering
     \includegraphics[width=0.75\linewidth]{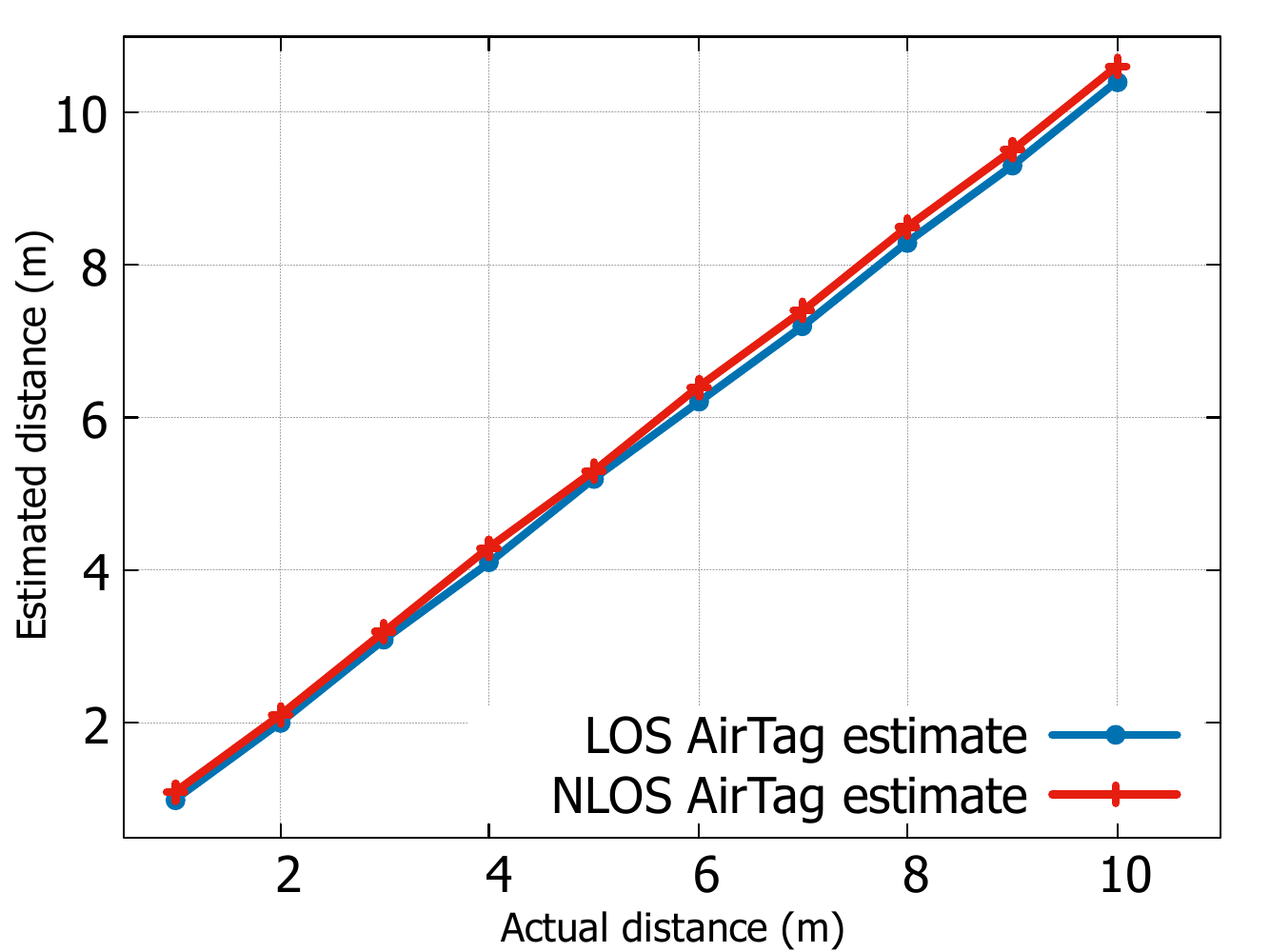}
    \caption{AirTags LOS \& NLOS experiments.}
    \label{fig:AirTag-LOS&NLOS-estimate}
    \end{minipage}
    \vspace{-1cm}
\end{figure}

\subsection{Feasibility Study}\label{sec:feasibility-study}
In this section, we show the results derived from experimental investigations aimed at assessing the feasibility of employing AirTags for localization purposes and emphasizing their associated benefits. The experimental data was gathered within an office floor of a campus building, as comprehensively elucidated in Section~\ref{sec:system-evaluation}. We applied experiments to benchmark the estimated distance and angle. To evaluate the accuracy of AirTags, we placed several AirTags at known locations and measured the distance and angle between the user’s device and each AirTag under varying conditions, including both \ac{LOS} and \ac{NLOS} scenarios. In LOS scenarios, where there is a clear path between the device and AirTag, and \ac{NLOS} scenarios, where obstacles such as walls obstruct the signal, we observed only minimal degradation in ranging and angular accuracy. Figure~\ref{fig:AirTag-LOS&NLOS-estimate} presents the results, showing that the AirTag’s \ac{UWB}-based measurements remained consistent and close to the actual values, even in \ac{NLOS} conditions. These findings confirm that AirTags provide highly reliable readings, achieving centimeter-level accuracy, making them ideal for indoor localization applications. The figure further illustrates that ranging errors remained within acceptable limits, reinforcing the practicality and precision of using AirTags for accurate indoor positioning.



In this section, we conducted an AirTag teardown to examine its internal components and confirm their suitability for indoor localization. Following this, we performed a feasibility study that validated the reliability of AirTags in various indoor environments. In the next section, we will discuss how \sys{} leverages AirTags to form a reliable system for precise indoor localization.

\section{System Overview}\label{sec:System-Overview}

\begin{figure}[!t]
    \begin{minipage}[b]{1\linewidth}
    \centering
     \includegraphics[width=0.75\linewidth]{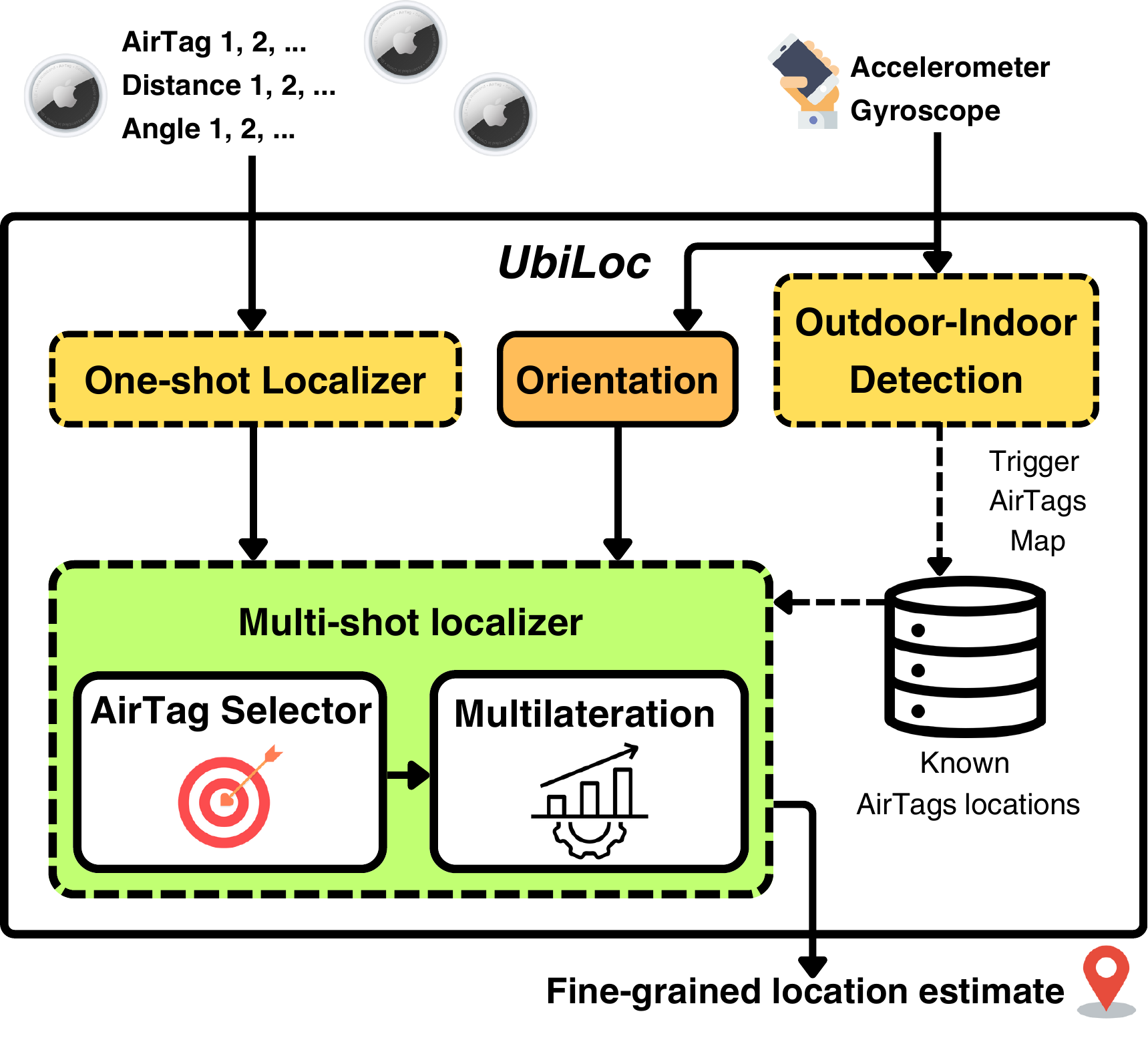}
    \caption{\sys{} architecture.}
    \label{fig:System-overview}
    \end{minipage}
    \vspace{-1cm}
\end{figure}

The architecture of the proposed system is illustrated in Figure~\ref{fig:System-overview}. It comprises four primary modules: \emph{Outdoor-Indoor Detection}, \emph{Pose Estimation}, \emph{One-shot Localizer}, and \emph{Multishot Localizer}. The following subsections provide an overview of each of these modules.

\subsection{Overview}
The \emph{Outdoor-Indoor Detection} module initially activates the system as the user moves from outdoor to indoor environments, ensuring both efficient operation and battery optimization. The system processes two main inputs: the first is a set of detected AirTags, accompanied by their respective ranging and angular measurements; the second is orientation data from the Compass/Gyroscope sensor in the user’s device, which tracks the user’s orientation relative to a reference AirTag.

The \emph{One-Shot Localizer} module combines these inputs to calculate the user’s relative position by converting the distance and angle data from the AirTags into Cartesian coordinates based on the user’s current pose using polar coordinate calculations. At the same time, the \emph{Multishot Localizer} module further refines these estimates, providing fine-grained accuracy. This integrated approach ensures consistent user localization across the testbed, even with varying AirTag densities.

In the balance of this section, we give the details of the different \sys{} modules.

\subsection{Outdoor-Indoor Detection}\label{Outdoor-Indoor-Detection}
The \emph{Outdoor-Indoor Detection} module mechanism plays a crucial role in the system by managing the shift between outdoor and indoor environments, helping to conserve battery life and activating the \emph{Multishot Localizer} only when necessary. As users move indoors, the GPS signal weakens or becomes unavailable. By detecting this transition, \sys{} can seamlessly switch to \emph{Multishot Localizer} for precise indoor localization. This detection process relies on monitoring GPS signal strength along with data from environmental context sensors on modern mobile devices. A significant drop in GPS signal, along with confirmation from accelerometer and light sensor data, triggers the activation of \emph{Multishot Localizer}. This method optimizes resource use while improving both the efficiency and accuracy of the localization process.

\subsection{One-shot Localizer}\label{subsec:One-shot_localizer}

At every user pose in the user trajectory, the cartesian location of AirTags is calculated from polar coordinates calculations for each AirTag in the area of interest. The range $\rho$ and angle $\theta$ acquired from each AirTag are used to calculate ($x, y$) location relative to the current pose location equation (\ref{eq:4}) \& (\ref{eq:5}) are applied \cite{zollo1999polar}. After, the cartesian location of the AirTag represents the anchor point for the \emph{Multishot Localizer} module. We compare the accuracy of multilateration techniques, emphasizing the selection of reliable AirTags in Section~\ref{sec:system-evaluation}.
\vspace{-0.5cm}

\begin{equation}\label{eq:4}
x = \rho\cdot \cos\left(\theta\right)
\vspace{-0.1cm}
\end{equation}
\begin{equation}\label{eq:5}
y = \rho\cdot \sin\left(\theta\right)
\end{equation}

To ensure accurate indoor localization despite the challenges posed by environmental factors, we employ a multi-sensor approach. Indoor environments with ferromagnetic materials and electrical objects create a turbulent magnetic field, which introduces significant noise that can negatively affect the accuracy of dead reckoning. To address this issue, we integrate the readings from both the gyroscope and the magnetic sensor. The gyroscope provides accurate and real-time changes in relative angles, while the magnetometer offers long-term stability. By leveraging the correlation between the two sensors, we can pinpoint moments when the compass reading is reliable. These reliable points are used as reference anchor points (AirTags), allowing \sys{} to measure the relative angle with the gyroscope until the next anchor points (AirTags) are detected \cite{mohssen2014s}.

\subsection{Multishot Localizer}\label{sec:Multishot-Localizer}

The user’s location is estimated at each pose using AirTags in the area, but without selecting the more reliable ones, drifting errors can occur. Noisy readings from distant AirTags, combined with the \emph{One-Shot Localizer} module, can worsen this drift. To mitigate this, \sys{} selects more reliable AirTags to improve localization and refine the user’s trajectory.

\begin{figure}[!t]
	\centerline{\includegraphics[width=0.4\textwidth]{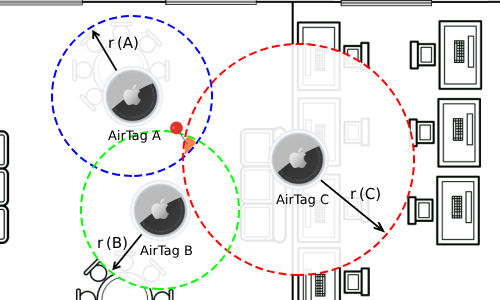}}
	\caption{In the context of 2D multi-lateration, the user's location can be determined by identifying the intersection area among three rings centered around reference points A, B, and C. Each reference point has its own ring, distinguished by varying widths proportional to the standard deviation observed in the ranging measurements.}
	\label{fig:multi-lateration-concecpt-using-airtags}
 \vspace{-0.5cm}
\end{figure}

\subsubsection{The AirTag Selector}\label{subsubsec:airtag-selector}
\hfill\\
In accordance with the distribution of AirTags within the designated area, it is common to have multiple AirTags visible at a given location simultaneously. However, due to the presence of noise in the wireless propagation channel, some of the received AirTags' ranging measurements may contain outlier values, as shown in Section~\ref{sec:feasibility-study}. Consequently, the localization process can be improved by selecting a subset of the visible AirTags for further analysis.

To address this, the AirTag Selector module employs various heuristics to determine the most suitable AirTags for localization purposes. These heuristics encompass selecting the $k$ strongest AirTags, the $k$ closest AirTags based on ranging measurements, or the $k$ AirTags with the lowest variance in estimated range values. In Section \ref{sec:system-evaluation}, we conduct a comprehensive performance evaluation to compare the efficacy of these distinct heuristics.

\subsubsection{Multilateration}\label{subsec:multilateration}
\hfill\\
We use multilateration (\ref{eq:1}) \& (\ref{eq:2}) \& (\ref{eq:3}) as a key technique in our system to calculate the user’s position by measuring distances from multiple anchor points with known locations, such as AirTags. Multilateration works by determining the user’s location based on the time-of-flight or signal strength of the \ac{UWB} signals received from the AirTags. By using the known positions of these anchor points, the system can calculate the distance between each AirTag and the user’s device, and from there, it triangulates the exact position of the user. In our testbed, shown in Figure~\ref{fig:multi-lateration-concecpt-using-airtags}, we deploy AirTags at known locations throughout the environment to serve as anchor points. The system’s accuracy depends heavily on the precision of the ranging measurements and the reliability of the AirTags selected. To enhance this accuracy, we employ a selection technique that prioritizes the most reliable AirTags, reducing the impact of noisy or distant signals, which can introduce errors in localization. In Section~\ref{sec:system-evaluation}, we will compare the performance of different multilateration techniques and demonstrate the effectiveness of our AirTag selection process in minimizing localization errors, ensuring high precision in real-world environments.

\begin{equation}\label{eq:1}
(x - x_1)^2 + (y - y_1)^2 = d_1^2
\vspace{-0.4cm}
\end{equation}

\begin{equation}\label{eq:2}
(x - x_2)^2 + (y - y_2)^2 = d_2^2
\vspace{-0.4cm}
\end{equation}

\begin{equation}\label{eq:3}
(x - x_3)^2 + (y - y_3)^2 = d_3^2
\end{equation}

In Section~\ref{sec:system-evaluation}, we conduct a comprehensive performance evaluation to compare the effect of changing parameters and choosing techniques.


\section{Syetem Evaluation}\label{sec:system-evaluation}
This section provides an in-depth evaluation of the \sys{}’s performance across two different testbeds utilizing AirTags and iPhones. The evaluation begins with the effect of different parameters on the system’s overall performance. Finally, we perform a comparative analysis to evaluate the localization accuracy of the system, emphasizing the role of AirTag selection in improving accuracy.

\begin{figure}
 \centering
\subfloat[Testbed 1: Campus Building. \label{fig_layout_campus}]{%
  \includegraphics[width=0.49\linewidth]{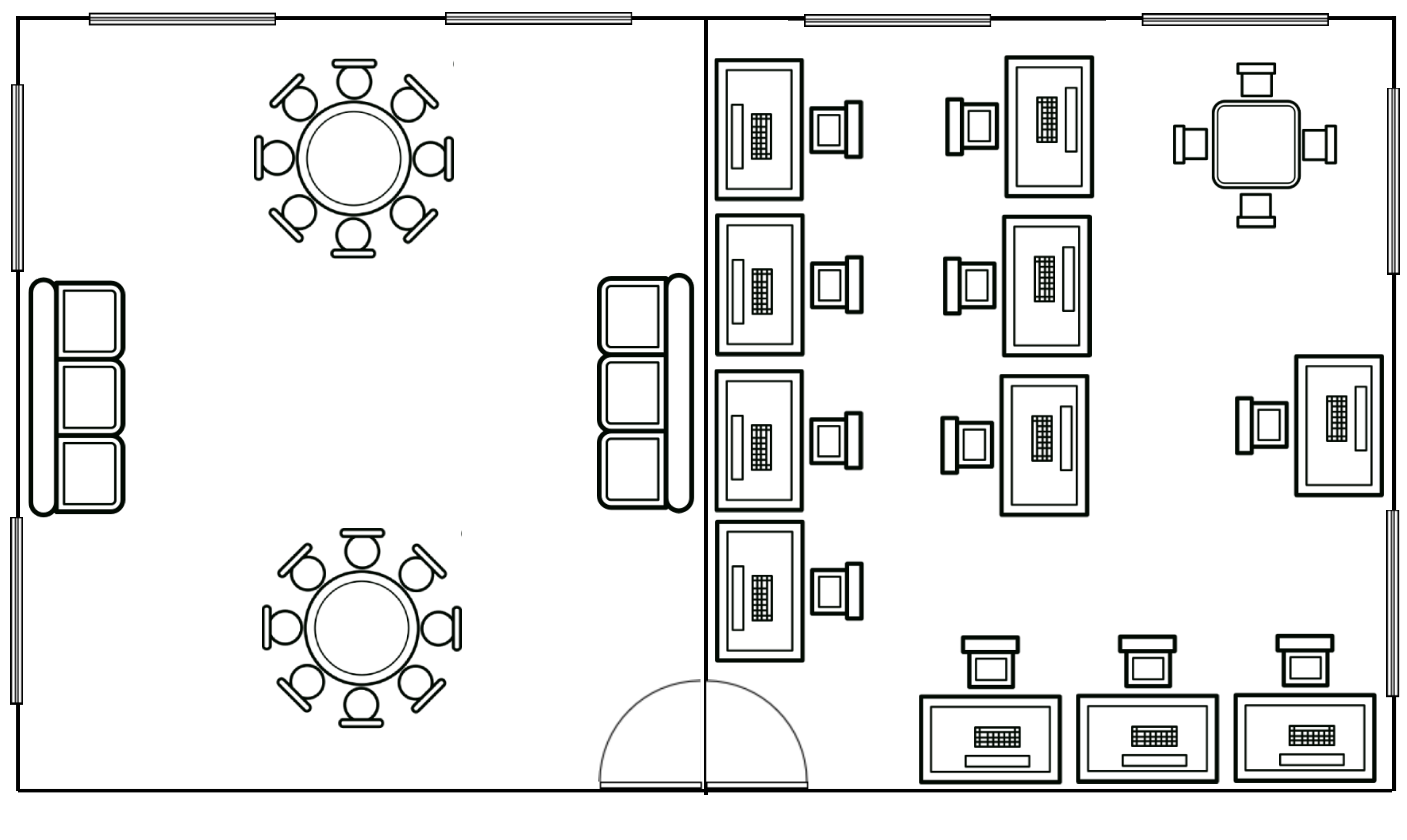}}
\hfill
\subfloat[Testbed 2: Apartment. \label{fig_layout_apartment}]{%
  \includegraphics[width=0.49\linewidth]{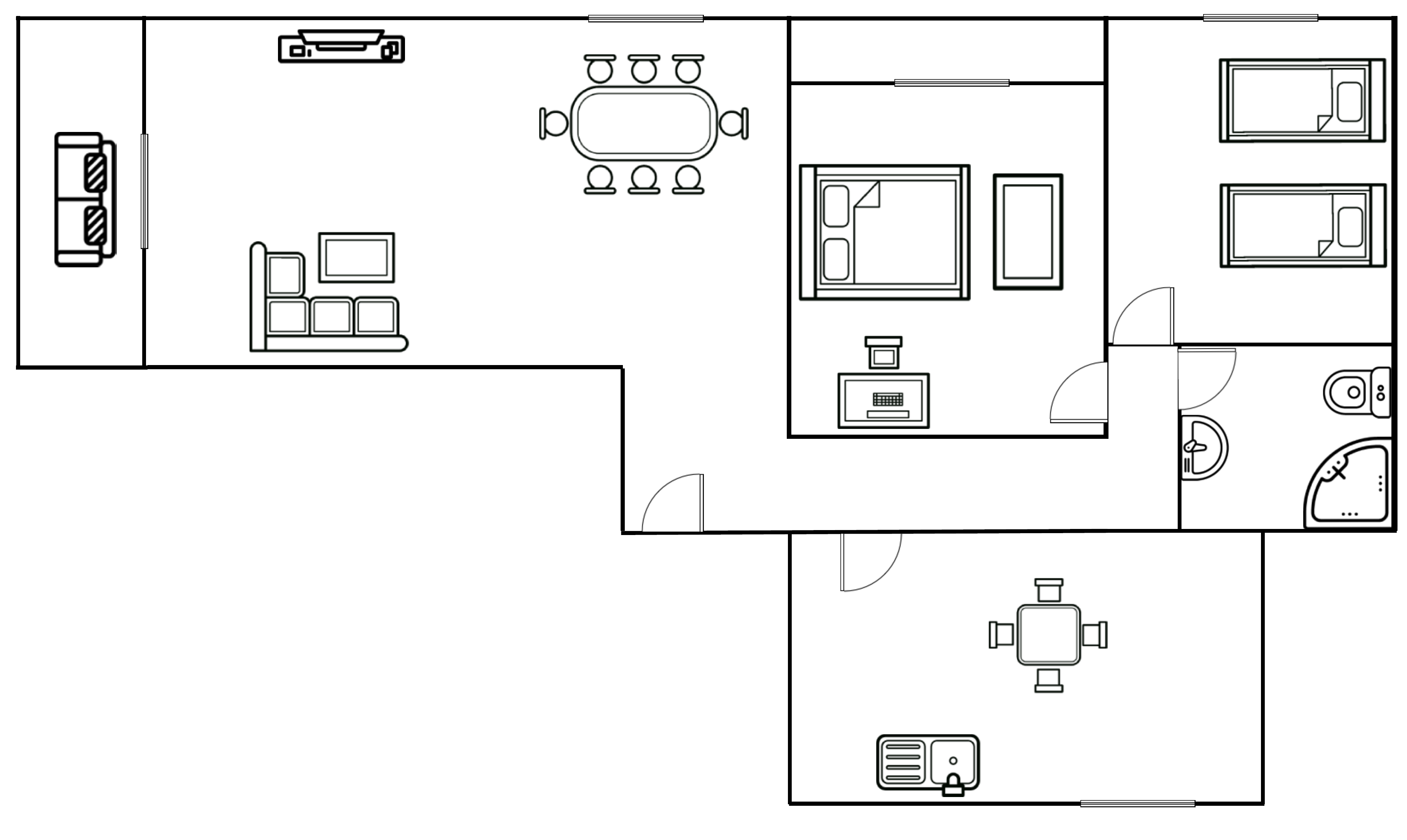}}
\caption{Floor plans of our two evaluation testbeds.}
\label{fig_layouts}
 \vspace{-0.65cm}
\end{figure}

\begin{table*}[!t]
    \caption{\sys{} Parameters and Their Default Values.}
    \label{table_testbeds_parameters}
    \centering
  \begin{tabular}
  {|p{12em}|p{19em}|p{15em}|}
    \hline
\textbf{Parameter} &\textbf{Range} &\textbf{Default Value}\\
    \hline
            \textit{Testbed} & Campus building, Apartment & Campus building\\
            \hline
            \textit{Device} & iPhone 11, 13 Pro, 13 Pro Max & iPhone 13 Pro\\
            \hline
            \textit{AirTag selection} & All, k-Nearest, k-Farthest, and k-Least variance & k-Nearest\\
            \hline
            \textit{Max. selected AirTags (k) } & [1:12] & 6\\
    \hline
  \end{tabular}
\end{table*}

\begin{figure*}
\minipage[t]{0.32\textwidth}
  \includegraphics[width=\linewidth]
  {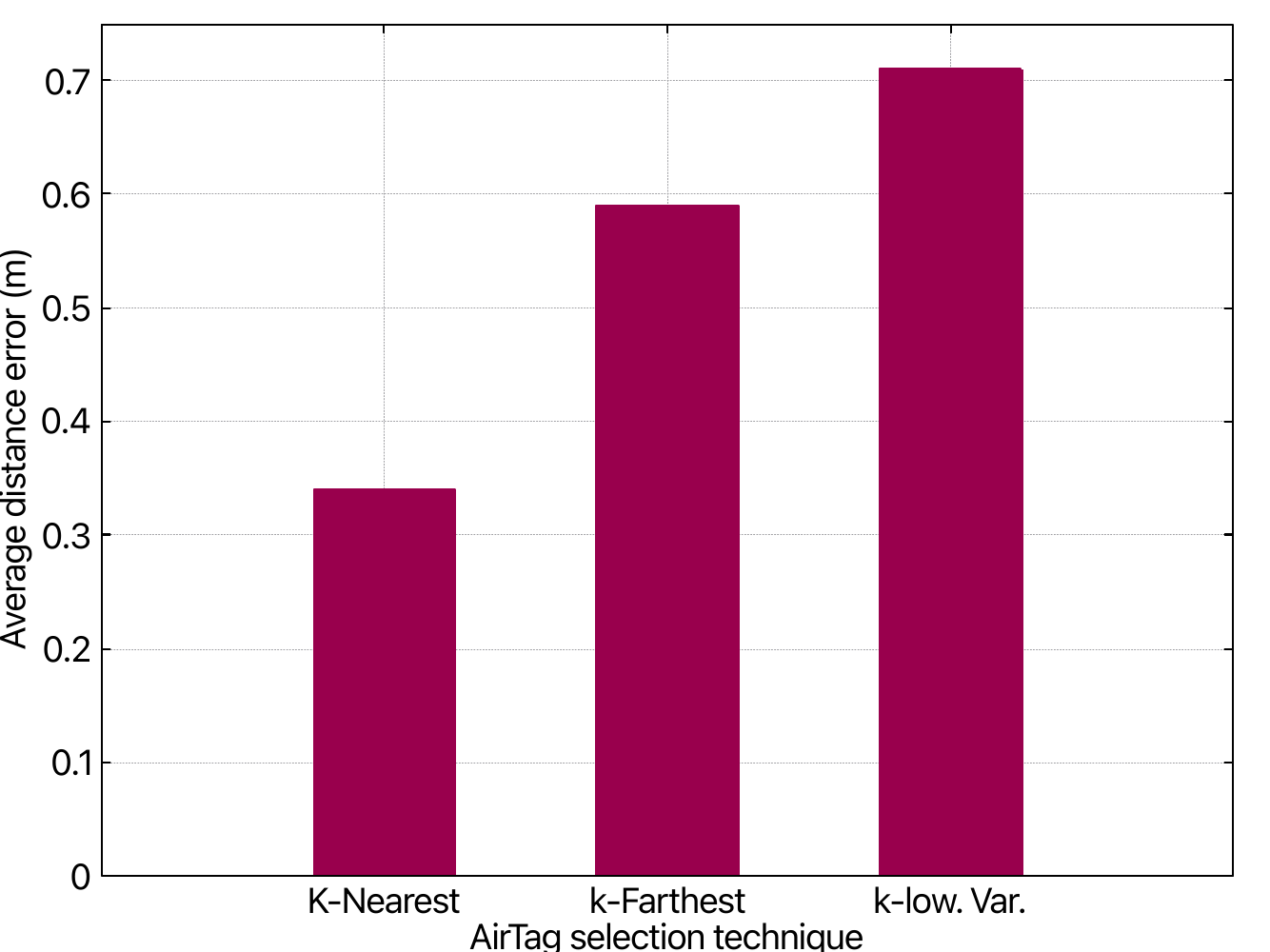}
  \vspace{-0.5cm}
  \caption{Effect of AirTag selection technique.}
  \label{fig_selection_eval}
\endminipage\hfill
\minipage[t]{0.32\textwidth}
  \includegraphics[width=\linewidth]
{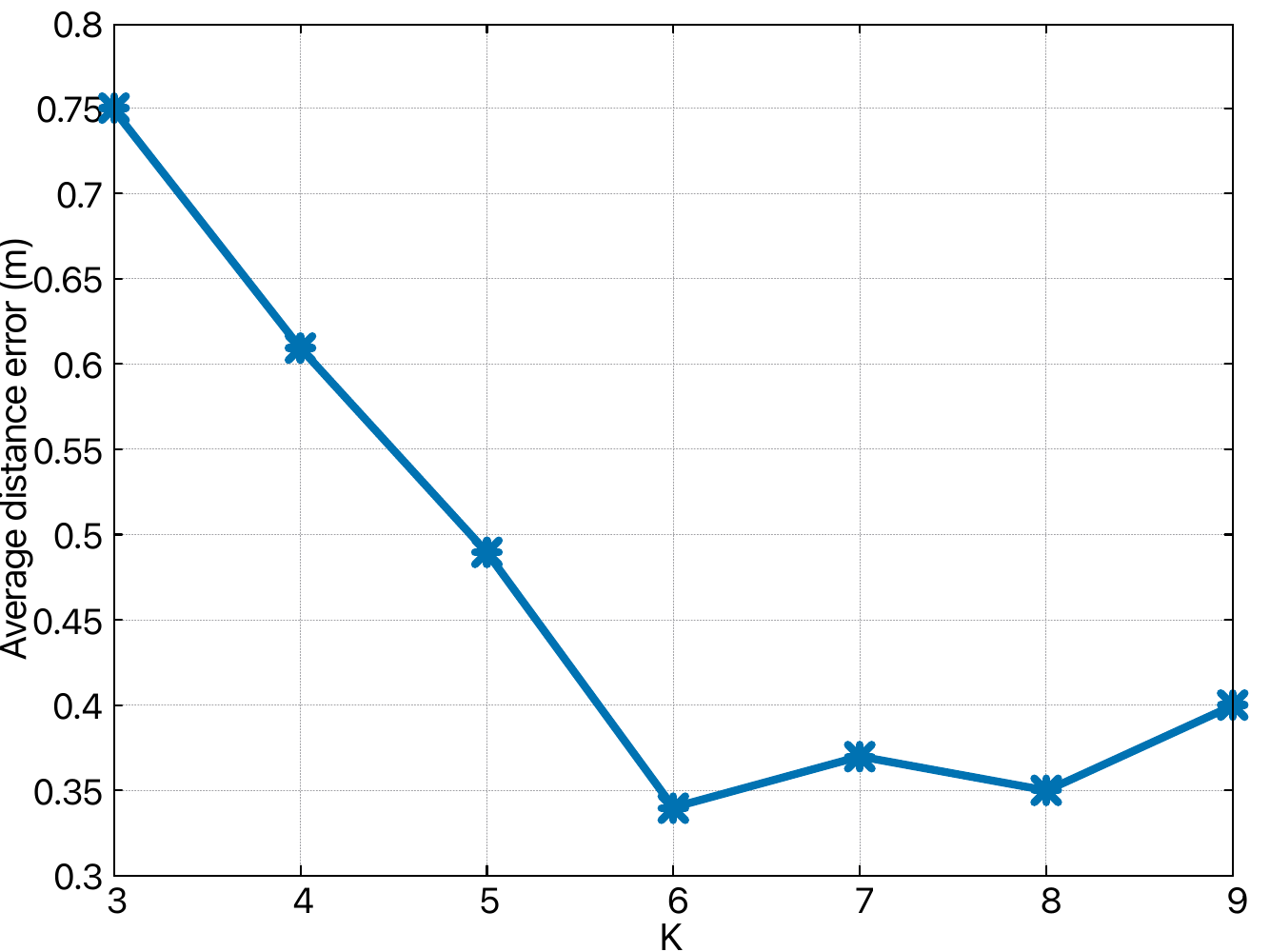}
\vspace{-0.5cm}
  \caption{Effect of the number of selected AirTags.}
  \label{fig_num_airtags_eval}
\endminipage\hfill
  \minipage[t]{0.32\textwidth}
  \includegraphics[width=\linewidth]{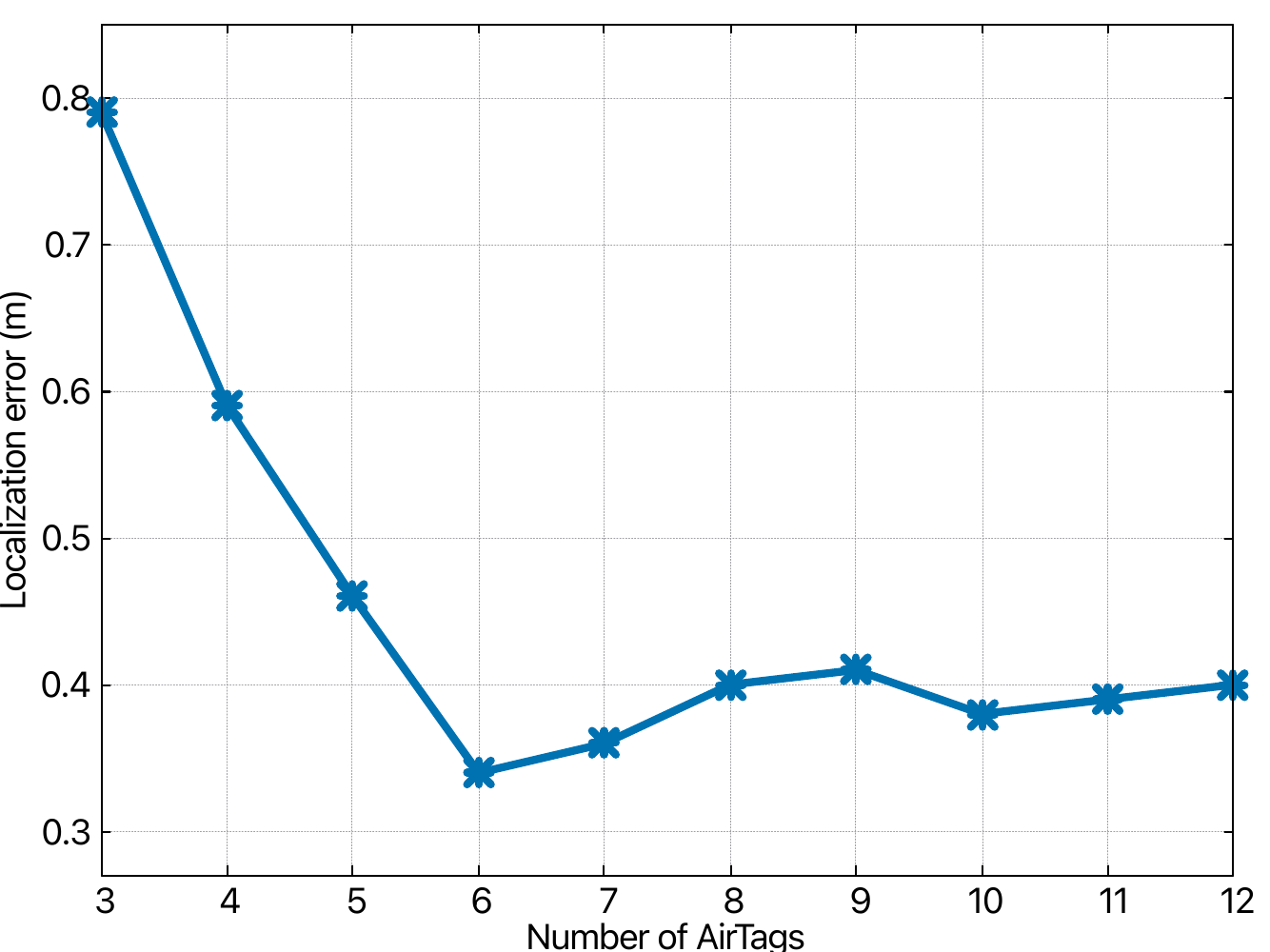}
  \vspace{-0.5cm}
  \caption{AirTags Density in Campus testbed.}
  \label{fig_AirTags_density}
\endminipage
\vspace{-0.7cm}
\end{figure*}

\begin{figure}
\centering
  \includegraphics[width=0.7\linewidth]
{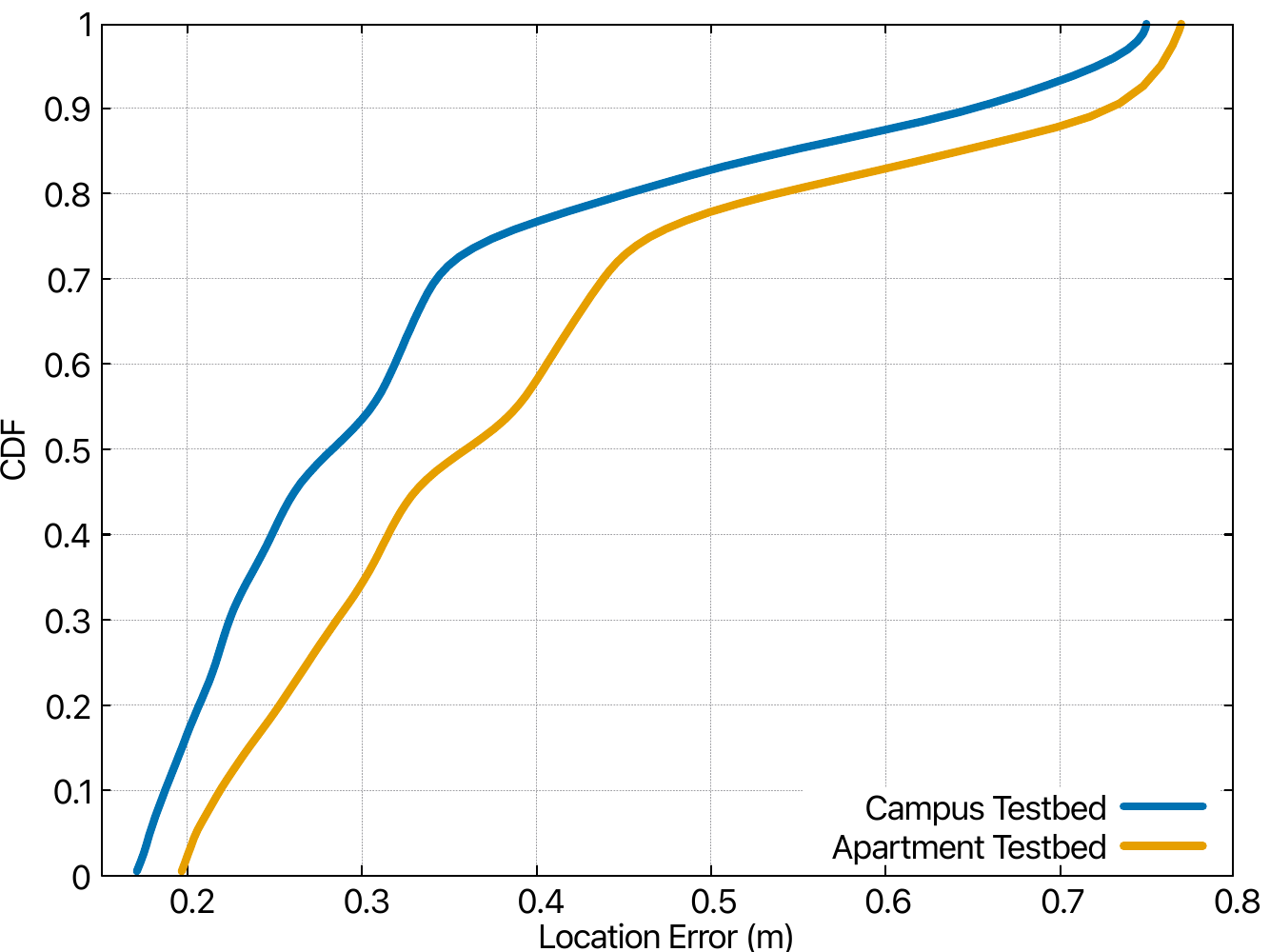}
  \caption{Localization error CDF in the Campus and Apartment Testbed.}
  \label{fig_CDF-user-location-campus-apartment}
  \vspace{-0.5cm}
\end{figure}

\subsection{Effect of Changing \sys{} Parameters}
In this section, we examine the impact of different \sys{} parameters on the user localization accuracy. Table~\ref{table_testbeds_parameters} lists the ranges of the parameters and their default values. For presentation clarity, we use the \textit{campus building} testbed to examine the effect of the parameters while we report the overall accuracy on both testbeds in Section~\ref{sec_eval_overall} using the default parameter values.

\subsubsection{AirTag Selection Technique}
Figure~\ref{fig_selection_eval} compares the average localization error for different AirTag selection techniques from Section~\ref{sec:Multishot-Localizer} with selecting all AirTags. By fixing the number of selected AirTags ($k$), the k-nearest method outperforms the others, as it selects AirTags with less noise, resulting in more accurate ranging measurements.

\subsubsection{Number of Selected AirTags}
Figure~\ref{fig_num_airtags_eval} shows that using the k-nearest AirTag selection technique, \sys{}’s localization accuracy saturates at $k = 6$, with no significant improvements beyond this point. This is due to the negative impact of farther AirTags on accuracy. Thus, we set $k = 6$ to achieve the best accuracy while improving computational efficiency.



\subsection{Overall \sys{} Performance}\label{sec_eval_overall}
In this section, we discuss the overall localization accuracy of \sys{} in the two testbeds using the default parameter values indicated in Table~\ref{table_testbeds_parameters}. Moreover, we study the accuracy of \sys{} over time to highlight the effectiveness of the AirTags as error resetting anchors. 


\subsubsection{\sys{} Overall Localization Accuracy}
Figure~\ref{fig_CDF-user-location-campus-apartment} shows the CDF of localization error for multilateration in the campus building and apartment testbeds. \sys{}’s median localization error is $26~cm$ and $31.5~cm$, respectively, only 130.8\% and 138\% worse than standard multilateration. \sys{} achieves these results by leveraging its optimized Multishot Localizer and effectively addressing challenges related to selecting reliable AirTags and handling inaccurate ranging measurements.


\subsubsection{Effect of AirTags Density}
Figure~\ref{fig_AirTags_density} illustrates the impact of AirTag density on UbiLoc’s localization accuracy. As the number of AirTags increases, the localization error decreases until the density reaches 6 AirTags, after which it saturates. This is due to improved signal availability and better spatial coverage at lower densities. However, beyond this point, adding more AirTags offers diminishing returns. This suggests that 6 AirTags is the optimal number to balance performance and system efficiency without overcrowding the environment.

\section{Related Work}\label{sec:Related-Work}
In this section, we discuss the prior work related to \sys{} and indoor localization systems.

Indoor localization systems can be broadly categorized based on the localization technique into fingerprinting-based systems \cite{bahl2000radar, youssef2005horus, shokry2017tale, hoang2018soft, rizk2019monodcell, hashem2020winar, abbas2019wideep, wang2015deepfi, ibrahim2018cnn, wu2019efficient} and ranging-based systems \cite{elbakly2016robust, pizarro2021AccurateUbiquitousLocalization, brunacci2023DevelopmentAnalysisUWB}.

\subsection{Fingerprinting-based Systems}
Fingerprinting-based indoor localization systems \cite{bahl2000radar, youssef2005horus, shokry2017tale, hoang2018soft, rizk2019monodcell, hashem2020winar, abbas2019wideep, wang2015deepfi, ibrahim2018cnn, wu2019efficient, rizk2020gain, rizk2019solocell, yang2012locating, wang2016csi} involve two main phases: an offline \textbf{calibration} phase and an online \textbf{prediction phase}. In the offline phase, the system builds a "fingerprint map" of the environment to map certain signal measurements, e.g., WiFi \ac{RSS} \cite{bahl2000radar, shokry2017tale, youssef2005horus, ibrahim2018cnn, fahmy2021monofi, ferris2007wifi, meng2011secure} or \ac{RTT} \cite{hashem2020winar, mohsen2023locfree}, to the locations in which they are obtained and accordingly builds a location inference model based on this fingerprint map. In the online phase, the inference model prepared offline is used to predict the most probable user location based on real-time signal measurements. 
The key limitation of fingerprinting-based systems is the labor-intensive offline calibration, which must be repeated in every new environment to build the fingerprinting map, making it especially burdensome for deep learning-based solutions \cite{rizk2018cellindeep, wang2015deepfi, abbas2019wideep, ibrahim2018cnn, rizk2019monodcell}. Some systems propose approaches like crowdsourcing to decrease the calibration overhead \cite{shokry2017tale}; nonetheless, they either require extra infrastructure costs or result in decreased accuracy due to data collection and labeling errors. Moreover, typical fingerprinting-based systems cannot achieve cm-level localization accuracy using commercial user devices \cite{wang2015deepfi, azizyan2009surroundsense}.

In contrast, \sys{} proposes a \textbf{calibration-free} localization system that does not require any prior knowledge or data collection overhead. Furthermore, as quantified in the evaluation section, \sys{} is able to achieve \textbf{cm-level} localization accuracies using the increasingly available commercial AirTag devices.

\subsection{Ranging-based Systems}
Ranging-based systems \cite{elbakly2016robust, pizarro2021AccurateUbiquitousLocalization, brunacci2023DevelopmentAnalysisUWB, rizk2023laser} leverage ranging measurements to estimate the distance and/or angle from multiple transmitters to solve the localization problem geometrically depending on \textbf{known} and \textbf{fixed} transmitter locations. Although ranging-based indoor localization systems require much fewer calibration efforts compared to fingerprinting, they are less common since they either require special hardware to get timing and direction measurements for ranging \cite{rizk2020omnicells, pizarro2021AccurateUbiquitousLocalization, brunacci2023DevelopmentAnalysisUWB} or result in relatively low accuracies when relying instead on the ubiquitous but noisy \ac{RSS} measurements \cite{elbakly2016robust}.

On the other hand, \sys{} leverages \ac{UWB}-based ranging based on timing and direction measurements using \textbf{commercial} multi-antenna AirTags, providing \textbf{cm-level} localization accuracy. This makes \sys{} a \textit{ubiquitous}, \textit{highly-accurate}, and \textit{calibration-free} indoor localization system   

In contrast, \sys{} relies on \textit{commercial} user phones and \textit{increasingly available} AirTag devices, eliminating the need for special hardware or physical anchors in the environment. This makes \sys{} a more \textit{ubiquitous} and \textit{practically deployable} solution compared to existing systems. Additionally, \sys{} achieves \textit{cm-level} localization accuracy by addressing several technical challenges, such as managing ranging errors through novel AirTag selection mechanisms and optimizing parameter selection.

\vspace{-0.4cm}

\section{Conclusion}\label{sec:conclusion}
In this paper, we presented \sys{}, a calibration-free indoor localization system that leverages commercially available AirTags and user phones. We outlined the architecture and mechanisms of \sys{}, demonstrating how it achieves precise user localization by combining multilateration with novel AirTag selection and error management techniques. \sys{} was evaluated in two real-world test environments: a campus building and an apartment. Experimental results showed that \sys{} consistently achieves centimeter-level accuracy, with median localization errors of $26~cm$ in the campus testbed and $31.5~cm$ in the apartment setting. Compared to state-of-the-art methods such as Wi-Fi \ac{RTT} and traditional multilateration, which exhibit localization errors ranging from $0.1$ to $1~meter$ and $0.5$ to $2~meters$ respectively, \sys{} demonstrated superior accuracy. These findings highlight \sys{} as the first cm-level indoor localization system utilizing AirTags without the need for calibration, offering a highly practical and deployable solution. For future work, we plan to explore integrating additional sensing technologies, such as light and sound sensors, to further enhance \sys{}’s performance and versatility.

\bibliographystyle{IEEEtran}
\bibliography{references}
\end{document}